\documentclass[preprint,showpacs]{revtex4}
\usepackage{epsfig}

\begin{document}
\preprint{}

\title{Searching Low-Energy Conformations of  Two  Elastin Sequences }

\author{ Handan Ark{\i}n} 
\email{handan@hacettepe.edu.tr}
\affiliation{Department of Physics, Hacettepe University, Beytepe 06532,
 Ankara, Turkey }

\begin{abstract}
\noindent The three-dimensional structures of two common repeat motifs 
Val$^1$-Pro$^2$-Gly$^3$-Val$^4$-Gly$^5$ and 
Val$^1$-Gly$^2$-Val$^3$-Pro$^4$-Gly$^5$-Val$^6$-Gly$^7$-Val$^8$-Pro$^9$ 
of tropoelastin are investigated by using
the multicanonical simulation procedure. By minimizing the energy 
structures along the  trajectory the thermodynamically most stable
low-energy microstates of the molecule are determined. The structural
predictions are in good agreement with X-ray diffraction experiments.


\end{abstract}
\maketitle


\noindent Determination of the folded structure of biological macromolecules
such as polypeptides and proteins is an important goal in structural biology.
Because the three-dimensional structure gives their biological
 activity ~\cite{CHEMREV}.
The atomic interactions of a protein are commonly modeled by an empirical
potential energy function, which typically leads to a complex energy
landscape consisting of a tremendous number of  local minima.

\bigskip

\noindent The major difficulty in conventional protein simulations
such as the Metropolis Monte Carlo method or molecular dynamics lies in the 
fact that simulations are not effective at temperatures of experimental
interest because the system becomes trapped in one of a huge number
of energy local minima. The development of novel global optimization
algorithms  for the protein folding problem is still an active area 
of research. 
One way to overcome this problem is to perform simulation in a 
generalized-ensemble~\cite{HaOk99rev,Ok00rev} where each state is 
weighted by non-Boltzmann probability
weight factor, so that a flat histogram in potential energy space 
may be realized. This allows the simulation to escape from any
energy barrier and to sample much wider conformational space than conventional
methods. One of the well-known, powerful generalized-ensemble methods
is the Multicanonical algorithm (MUCA)~\cite{MUCA}.
The trapping problem can be alleviated by the multicanonical
simulation (MUCA) method.  In  one simulation all the temperature 
range is determined and all thermodynamic quantities are calculated.

\bigskip

\noindent The problem of protein folding entails the study of a 
nontrivial dynamics along patways embedded in a rugged landscape.
 Therefore,
development of efficient methods for conformational search
is of central importance.
 Methods for searching energy landscape and 
 low-energy conformations
are proposed~\cite{On,Wille,HATC03}, energy landscape perspectives are 
investigated~\cite{HATC2003}. Such a goal can be achieved within 
the multicanonical ensemble approach. 

\bigskip

\noindent While the  MUCA ensemble is based on a probability
function in which the different energies are equally probable:

\begin{equation} \label{multie}
P^{\rm MU}(E) \sim n(E) w(E) = constant
\end{equation}
where $w(E)$'s are multicanonical
weight factors. Hence, a simulation with this weight factor, which has
no temperature dependence, generates a one-dimensional random walk
in the energy space, allowing itself to escape from getting trapped
in any energy local minimum.

\bigskip

 Re-weighting techniques
(see Ref.~\cite{FeSw88})
enable one to obtain Boltzmann averages of various
thermodynamic properties over a large range of temperatures.
The advantage of this  algorithm lies in the fact that it not only
alleviates the multiple-minima problem but also allows the calculation of
various thermodynamic quantities as functions of temperature from
 one simulation run. This demonstrates the superiority of the method.

\bigskip

\noindent To verify the coverage of the low-energy region by the MUCA sample,
 I also
minimized the energy of 
  conformations
of the trajectory and indeed recovered the global energy 
minimized structure and
other low energy minimized structures~\cite{Deltorphin}. 
 This suggests that MUCA can also
serve as a useful conformational search technique
for identifying the most stable wide microstates
of a peptide. 
Further for comparison the effectiveness of searching low-energy
conformations, 
a  conformational search was obtained with
 the Monte Carlo minimization (MCM)
method of Li and Scheraga~\cite{LI}.

\bigskip

\noindent  The simulated  molecule, tropoelastin, is  soluble precursor
protein of fibrous elastin and  contains in its sequence several
stretches of repeating oligopeptides. The main three repeating
 oligopeptides are a tetrapeptide Val$^1$-Pro$^2-$Gly$^3$-Gly$^4$ , a
pentapeptide Val$^1$-Pro$^2$-Gly$^3$-Val$^4$-Gly$^5$ and a nanopeptide
Val$^1$-Gly$^2$-Val$^3$-Pro$^4$-Gly$^5$-Val$^6$-Gly$^7$-Val$^8$-Pro$^9$.
 In previous works, X-ray diffraction data were used to determine,
at room temperature, the
crystal structure of a repeat pentapeptide of elastin~\cite{xray}.
  Nuclear magnetic resonance (NMR)
studies have shown that the repeated-oligopeptide segments of elastin are
composed of subunits that are conformationally equivalent within the
NMR time scale.  Several secondary structural elements
have been proposed as features of one or more of these repeated
peptide segments~\cite{nmr}.
 Alternative approaches such as computer molecular modeling
starting from amino acid sequences can contribute to a better understanding
of the three-dimensional structures of these repeating oligopeptides.
In this work,  I determine {\it all}  the thermodynamically
stable conformations populated by the molecule, not only the stable
structure at room temperature.

\bigskip

\noindent The  tropoelastin sequences  
 namely
 the pentapeptide  Val$^1$-Pro$^2$-Gly$^3$-Val$^4$-Gly$^5$
and a nanopeptide
Val$^1$-Gly$^2$-Val$^3$-Pro$^4$-Gly$^5$-Val$^6$-Gly$^7$-Val$^8$-Pro$^9$
 are modeled  here by the well-known potential 
energy function ECEPP/2~\cite{ECEPP}, which
is given by the sum of the electrostatic term, 12-6 Lennard-Jones term and
the hydrogen bond term for all pairs of atoms in the peptide together
with the torsion term for all torsion angles.
 The
peptide bond angles $\omega$  are
kept fixed at 180$^{\rm o}$, to their common value,
and therefore a conformation of
 Val$^1$-Gly$^2$-Val$^3$-Pro$^4$-Gly$^5$-Val$^6$-Gly$^7$-Gly$^8$-Gly$^9$ 
sequence  is  
defined solely by 28 degrees of freedom, the 16 backbone dihedral angles
$\phi$ and $\psi$ ( the $\phi$ angles for the  proline residues are 
fixed) and the 12 side chain
dihedral angles $\chi$.
We use the standard  dielectric constant $\epsilon$=2 of ECEPP.
This force field is implemented into the
 software package FANTOM~\cite{FANTOM}. 


\noindent First, I carried out two  canonical (i.e., constant $T$) 
MC simulations
at 50 K and 500 K, and  multicanonical test
runs which enabled us to determine the required energy ranges.
At each MUCA update step a trial conformation was obtained by
changing  one dihedral angle
at random within  the range [$-180^{\rm o};180^{\rm o}]$, followed by
the Metropolis test. The dihedral angles were always visited in
a predefined  order, going from first to last residue; a cycle of $N$ MC steps
($N$=28) is called a sweep.

\bigskip

\noindent The MUCA weights  were built recursively  during a
long {\it single} simulation
where the multicanonical parameters were re-calculated every
5000~sweeps and 200~times
which adds up to 1,000,000 sweeps total.
Several such simulations were carried out and the final MUCA weights
of the best simulation were used in the following
MUCA production run of another
1,000,000 sweeps. In all cases, each multicanonical simulation started
from completely {\it random} initial conformation. No {\it a-prori}
information about the groundstate conformation is used in simulations.

\bigskip

\noindent As pointed out in the Introduction, for peptides it is not only of
interest to obtain thermodynamic averages and fluctuations at different
temperatures, but also to find the most stable regions in the conformational
space, which allows to identify the most stable wide microstates.
  Proteins are expected to
populate low energy wide microstates  even at room temperature, while
peptides might also populate relatively higher energy microstates.
Therefore,  it is of interest to investigate the conformational
coverage provided by MUCA, in particular in the low energy 
region~\cite{Deltorphin}.
In order to classify the microstates according to the potential
wells they belong around thermodynamically stable different structures, each
conformation of the simulation data was subjected to energy minimization.

\bigskip

\begin{figure}[!h]
\psfig{figure=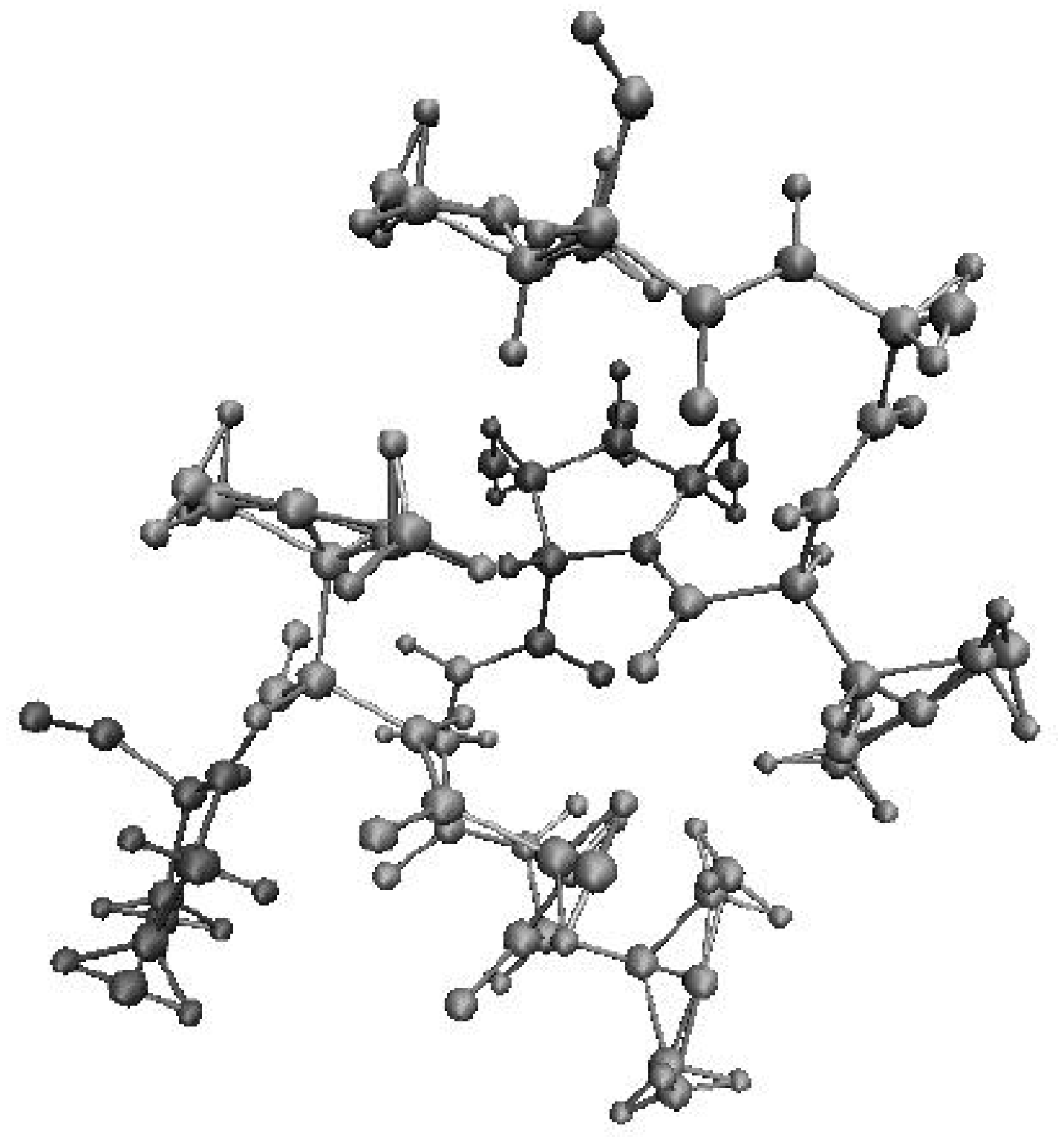,height=13.0cm,width=15cm}
\caption{ Structure of the conjectured GEM of the peptide sequence 
Val$^1$-Gly$^2$-Val$^3$-Pro$^4$-Gly$^5$-Val$^6$-Gly$^7$-Val$^8$-Pro$^9$
at energy $  E = - 17.94$ kcal/mol.  }
\label{3d}
\end{figure}

\noindent Following the methods proposed by
Meirovitch et al. ~\cite{MCM}, 
 the configurations generated in $10^6$
sweeps of the MUCA production run are minimized and  
 the minimized structures were
sorted according to a variance criterion where two structures are
considered to be different if at least one  dihedral angle differs by more
than 2$^\circ$. The lowest energy found (our suspected
GEM) is

\begin{equation} \label{E_GEM}
 E = - 17.94\,{\rm kcal/mol}\
\end{equation}
\noindent and its conformation is depicted in Fig~\ref{3d}. The number of 
structures found in energy bins of $ 0.5$ kcal/mol above $  E = - 17.94$
kcal/mol appear in Table~\ref{xx}. Only results of $10^5$ sweeps are presented
in the Table.
 Here we compare the conformational
searching of the low-energy region to
that obtained with MCM~\cite{LI}.
 For bins 7-12 the number for MCM 
is smaller than MUCA and for bins 1-4 the results of the two 
methods are  close which means that a very good coverage of the lowest 
energy bin is provided by MUCA.
From the table it is obvious that MUCA covers a large range of energies
in an approximately homogeneous way.

\bigskip

\begin{figure}[!h]
\centerline{\hbox{
\psfig{figure=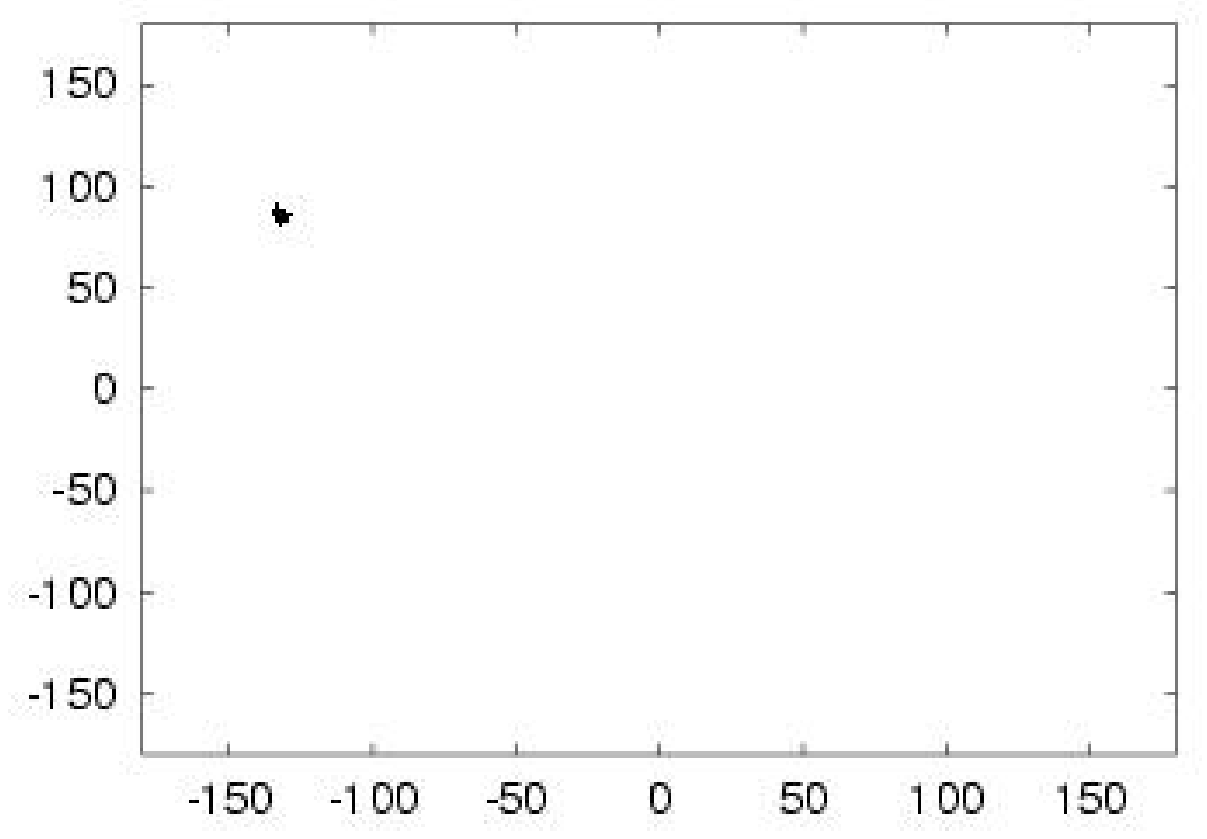,height=4.5cm,width=5cm}
\psfig{figure=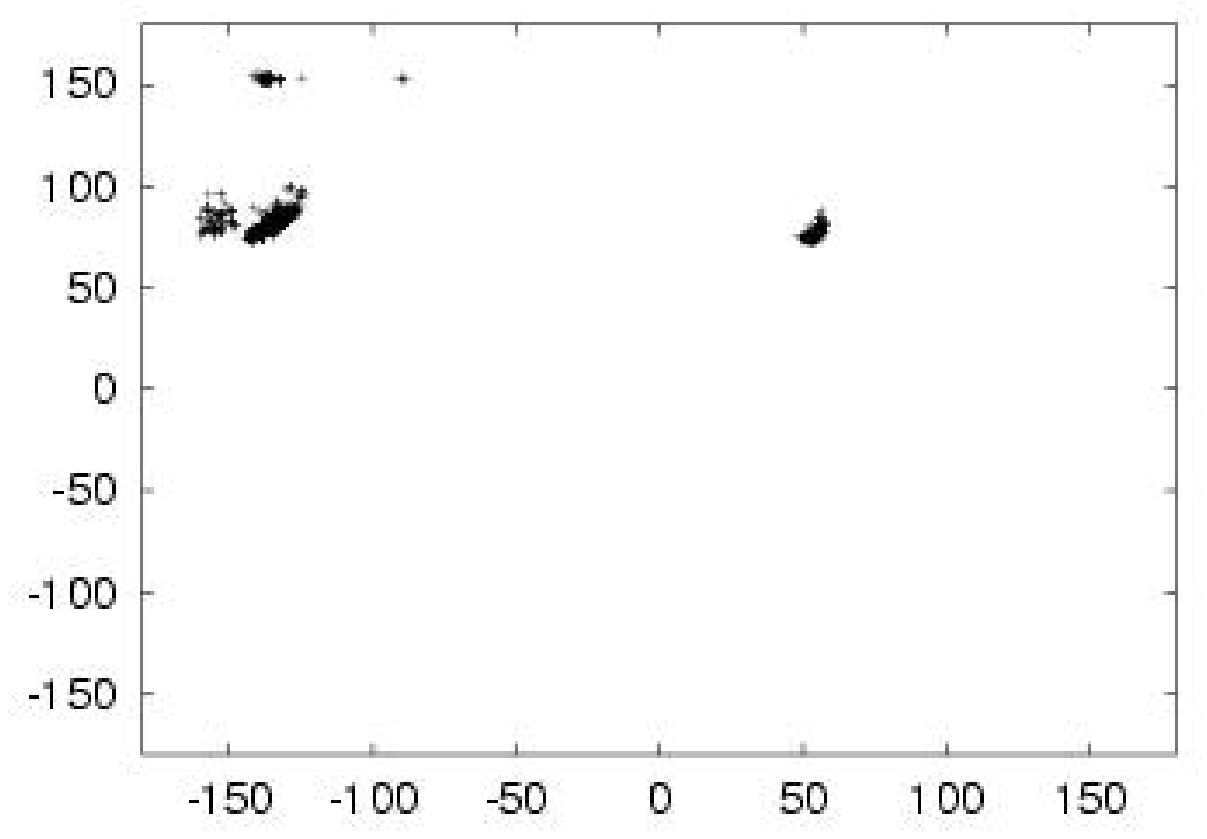,height=4.5cm,width=5cm} }}
\caption{Ramachandran plots of each residue (from top to bottom) 
Val$^1$-Gly$^2$-Val$^3$-Pro$^4$-Gly$^5$-Val$^6$-Gly$^7$-Val$^8$-Pro$^9$ ( 
except proline residues).
 The abcsisa is the angle $\phi$ and the ordinate is $\psi$. The angles
are in the range range of [$-180^{\rm o};180^{\rm o}]$. The first 
column shows the GEM and the conformations of the lowest energy bin 
above the GEM, the middle column shows conformations in the temperature 
range $(130-140)\,$K and the last column conformations in the temperatures 
range $(290-300)\,$K.}
\label{rama}
\end{figure}

\begin{figure}[!h]
\centerline{\hbox{
\psfig{figure=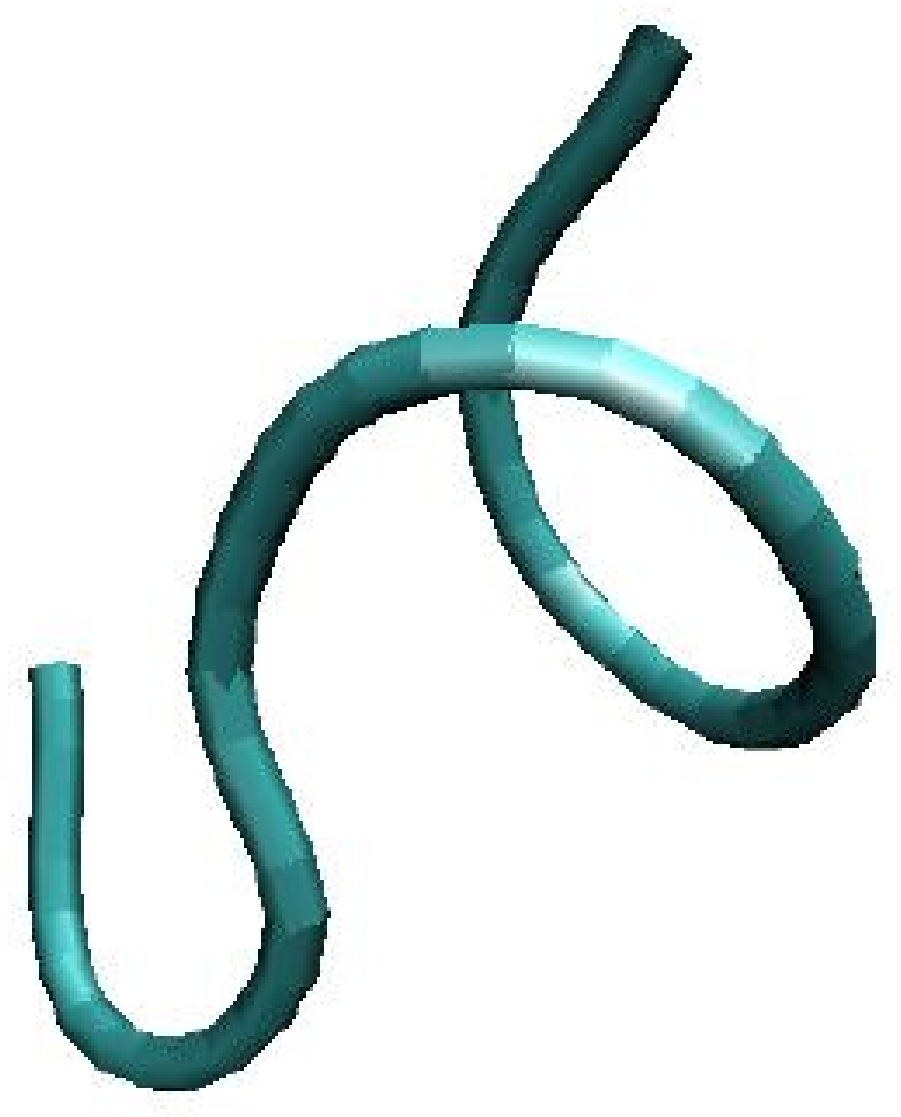,height=5.0cm,width=5cm}
\psfig{figure=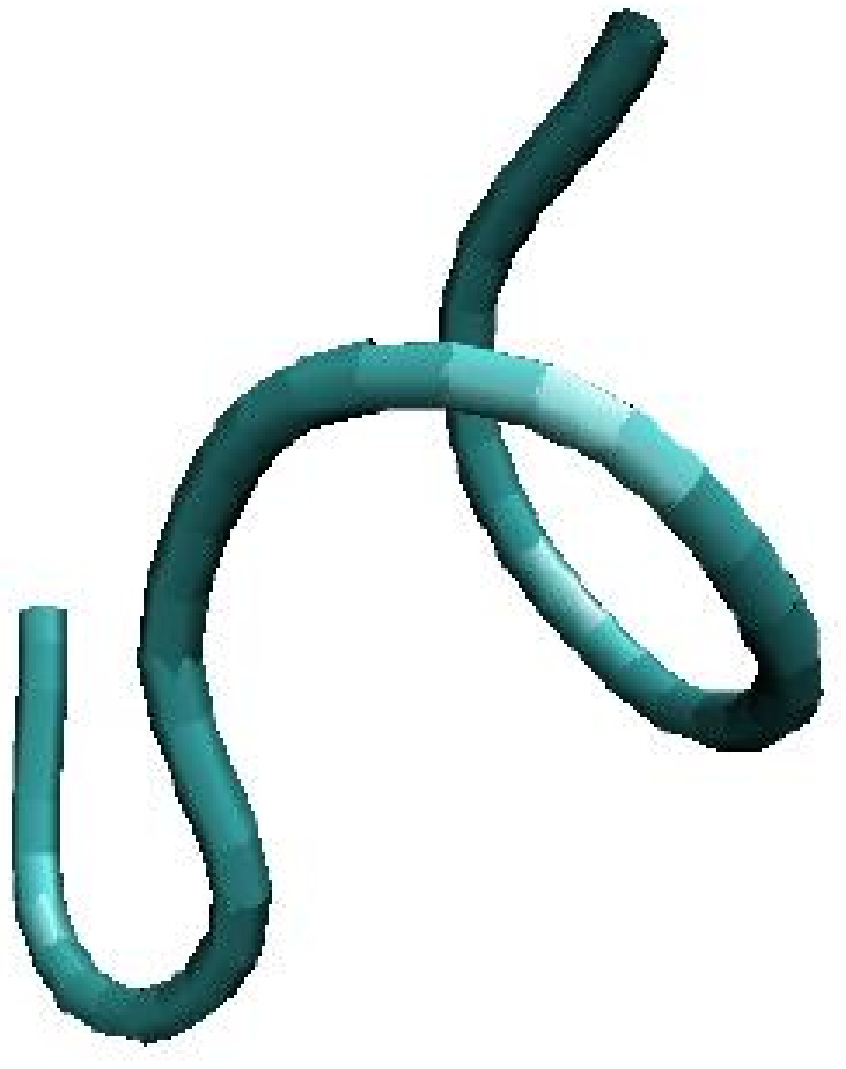,height=5.0cm,width=5cm}  
\psfig{figure=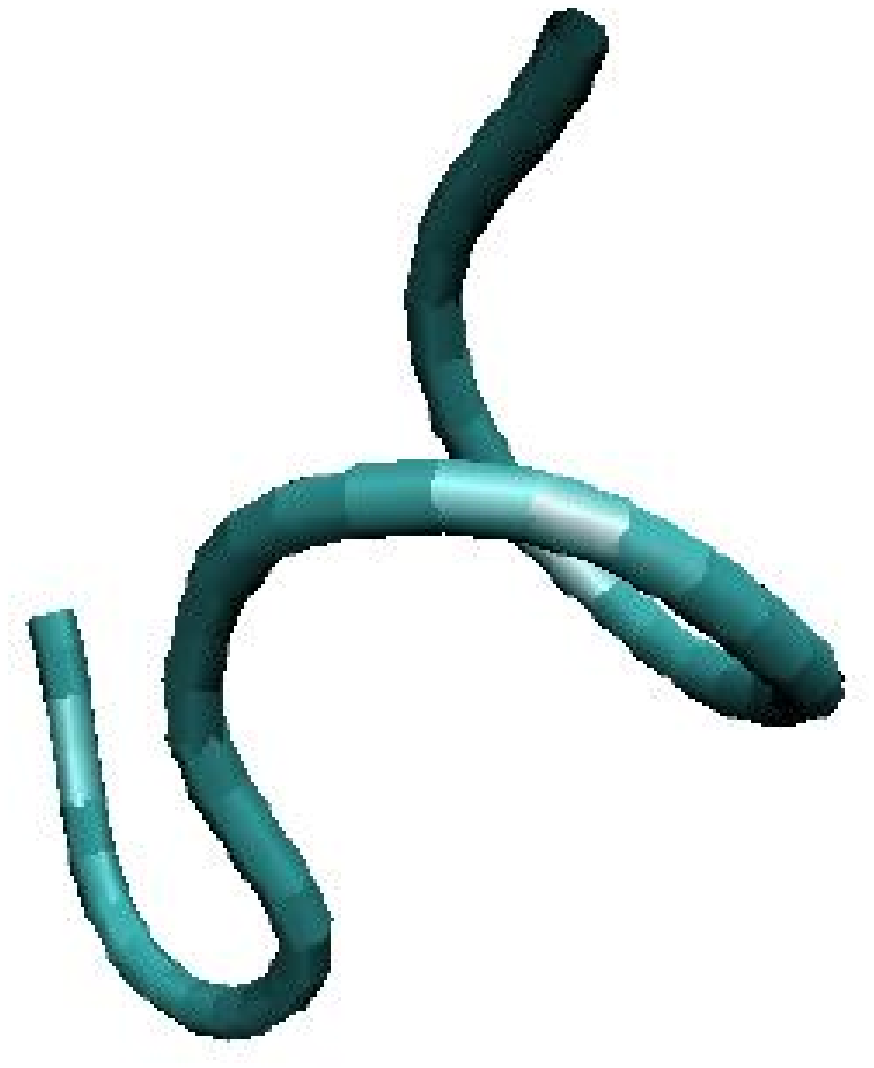,height=5.0cm,width=5cm} }}
\caption{Some  typical low-energy conformations with energies below
 $  E = - 17.45 $ kcal/mol ( first bin in Table~\ref{xx} ) of sequence
 Val$^1$-Gly$^2$-Val$^3$-Pro$^4$-Gly$^5$-Val$^6$-Gly$^7$-Val$^8$-Pro$^9$. }
\label{typical}
\end{figure}

\noindent The superiority of the multicanonical approach lies in
the fact that MUCA provides the sampling of conformations at all temperatures 
 from one simulation run, therefore enables one to study thermodynamics
of the system under consideration. The distribution of backbone 
dihedral angles
were analyzed and the Ramachandran
plots were prepared for each residue of conformations.  
In  Fig~\ref{rama}  the Ramachandran
plots for the nanopeptide sequence (except proline residues) of typical 
structures for three temperature  ranges are presented. The first set of
seven plots ( for the seven amino acid residues except prolines.)
is for GEM structure and structures pertaining to the lowest energy
bin above the GEM ( bin 1 in Table~\ref{xx} ). The second and
third sets shows the most probable structures for the two temperature
regions, 130 - 140 K and 290 - 300 K, respectively.
One can easily see that  the sample points in the set of plots at higher
temperatures ( the right column) are more scattered, but each plot contains 
the corresponding low temperature partner ( the plot at the left column)
as a subset. This means that these two groups  are in the same well-defined
valley in the energy space.
Therefore it becomes possible to track a chosen conformation at room 
temperature goes to which one of the lowest energy states as the temperature
lowered.
 Among the 1096 
conformations ( first bin in Table~\ref{xx} ), there are some typical
different backbone features to those of the global minimum fold. 
 The backbones  
of the  typical conformations   
with the lowest energies, are presented in Fig~\ref{typical}.
One can see this difference from the Ramachandran plots as well.
The backbone differences occur mainly in the terminal parts
 that are expected to be the most flexible.  Other 
differences comes from the side chain dihedral angles.  
 The scattering of points in Ramachandran plots increase  significantly for $
T = $ 290 - 300 K ( third column in Fig~\ref{rama}), suggesting 
that the sequences modeled with the ECEPP potential
will populate at least several wide microstates. 
Analyzing the Ramachandran plots, the structures at $T =  $ 290 - 300 K
consists of $\beta$-turns in Val$^1$-Gly$^2$-Val$^3$,  Pro$^4$-Gly$^5$ and 
 Val$^6$-Gly$^7$-Val$^8$ bridges.
 These  results are in agreement with previous studies, both
experimental~\cite{exp} and simulation~\cite{sim} approaches,
which suggest that the main secondary structure in elastin 
is short $\beta$-turns.  Circular dichroism (CD)
and  NMR measurements gave evidence  of flexible $\beta$-turns as the 
dominant structural feature~\cite{tam, Urry}.

\bigskip

\noindent  In addition, the plots of Val$^3$-Pro$^4$-Gly$^5$-Val$^6$-Gly$^7$
part of the nanopeptide are very similar   to the Ramachandran
 plots of the pentapeptide
( Val$^1$-Pro$^2$-Gly$^3$-Val$^4$-Gly$^5$ ), which is also simulated 
independently ( plots are not presented.). 

\bigskip

\begin{figure}[!b]
\hspace*{0.2cm}
\psfig{figure=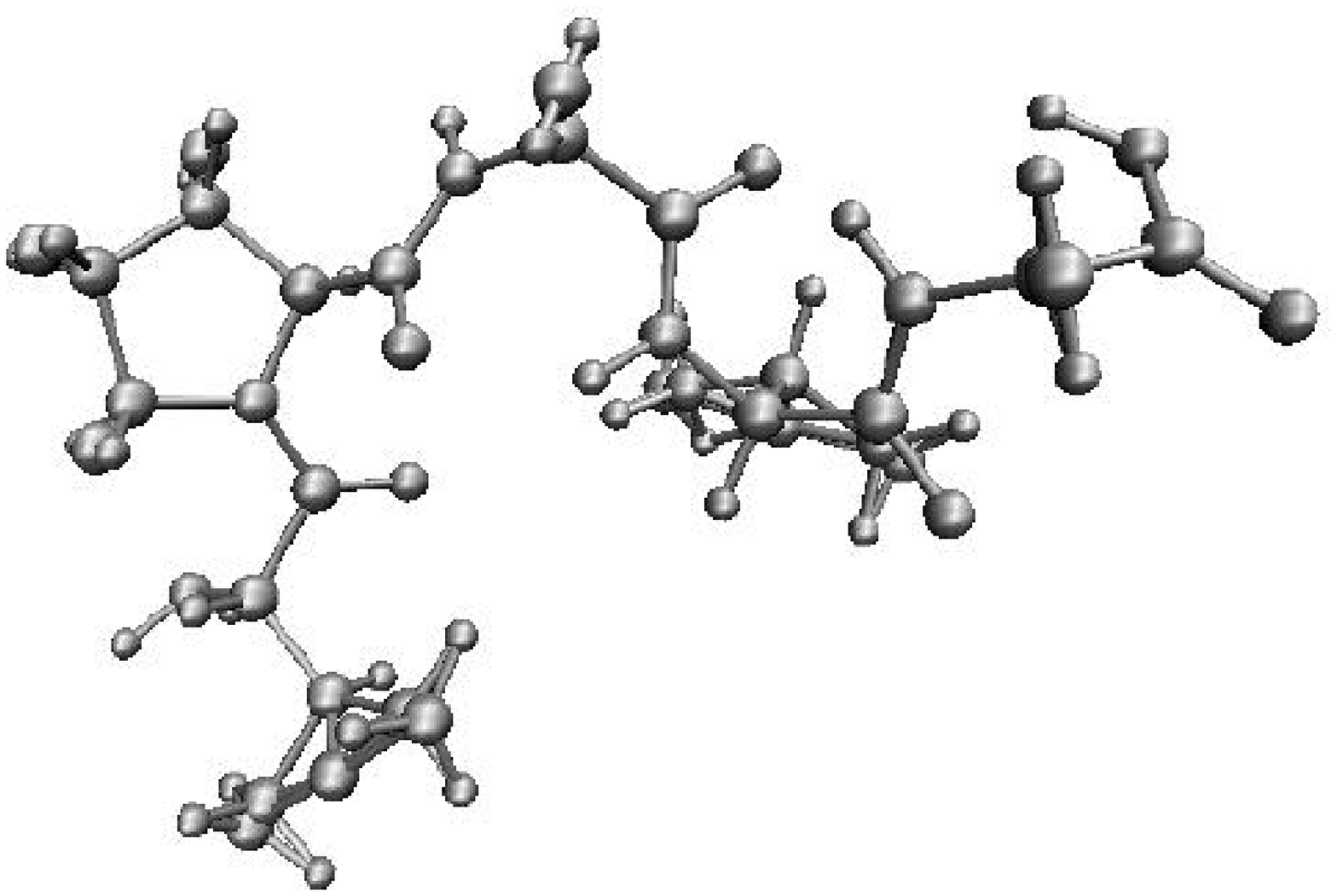,height=9.0cm,width=12cm}
\caption{a) The conformation of pentapeptide sequence obtained 
with X-ray diffraction experiments. The data is from the paper: W. J. 
Cook et al., J. Am. Chem. Soc., 102, 5502 (1980). b) The conformation
of the pentapeptide obtained at room temperature by multicanonical simulation.}
\label{oda}
\end{figure}

\begin{figure}[!h]
\psfig{figure=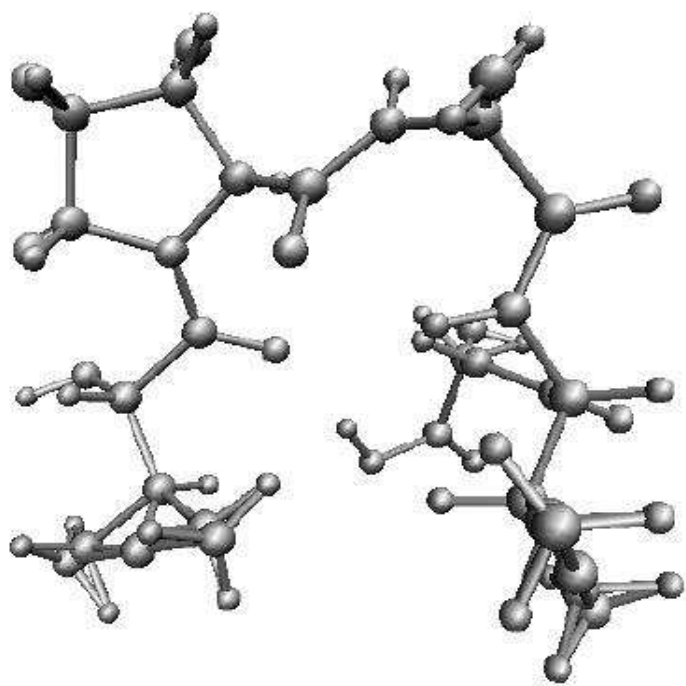,height=10.0cm,width=12cm}
\caption{Structure of the conjectured GEM of the pentapeptide sequence
Val$^1$-Pro$^2$-Gly$^3$-Val$^4$-Gly$^5$ with energy $  E = - 4.40 $ kcal/mol .}
\label{3d2}
\end{figure}

\noindent To compare the experimental findings and the multicanonical 
simulation results,
the stable conformation of VPGVG ( in one letter code) sequence
at room temperature and the X-ray data are both shown in
Fig~\ref{oda}.
The crystal structure obtained with X-ray diffraction experiments are very
close to  the thermodynamically stable conformation  at 
room temperature which is found
by multicanonical simulation.   
 The overlap parameter value ( see Ref.~\cite{HATC2003} for the
definition of the overlap parameter)  between the two
structures are found 0.94 which means that the two conformations 
are very close
to each other. In  
Fig~\ref{3d2} the global minimum energy conformation of VPGVG
sequence is given. 
By inspecting the Ramachandran plots, one can see that the global 
minimum conformation of Fig~\ref{3d2} and the conformation at
 room temperature shown in Fig~\ref{oda}, are in the same energy valley and 
 the molecule  at room temperature is 
 folded almost to  its native structure at low temperature.

\bigskip

\noindent To summarize, two common repeat motifs of tropoelastin have been
simulated by using the multicanonical approach and demonstrated that 
a very good
coverage of the conformational space especially in the low energy region is
achieved. Sampling of conformations pertaining
different microstates over wide range of temperatures, the most stable
 low energy microstates and the global energy minimum are obtained by 
multicanonical simulations.
 The results confirm the experimental findings of X-ray 
studies~\cite{xray} and CD studies~\cite{tam}.
Condidering computer time, MUCA simulation  required a 9 hour production run
of 1 million sweep 
for nanopeptide on a DEC-Alpha 433 workstation.

\clearpage
\pagebreak

\pagebreak

\begin{figure}[!h]
\centerline{\hbox{
\psfig{figure=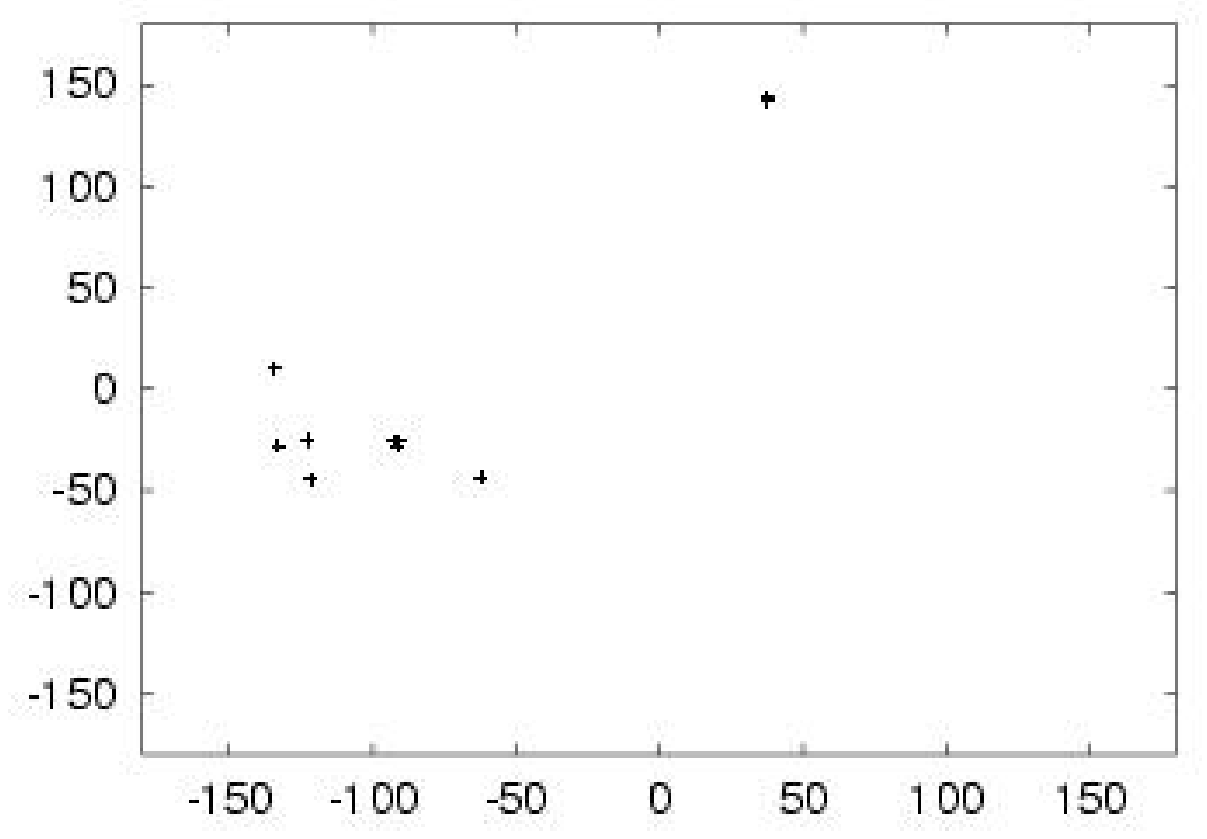,height=4.5cm,width=5cm}
\psfig{figure=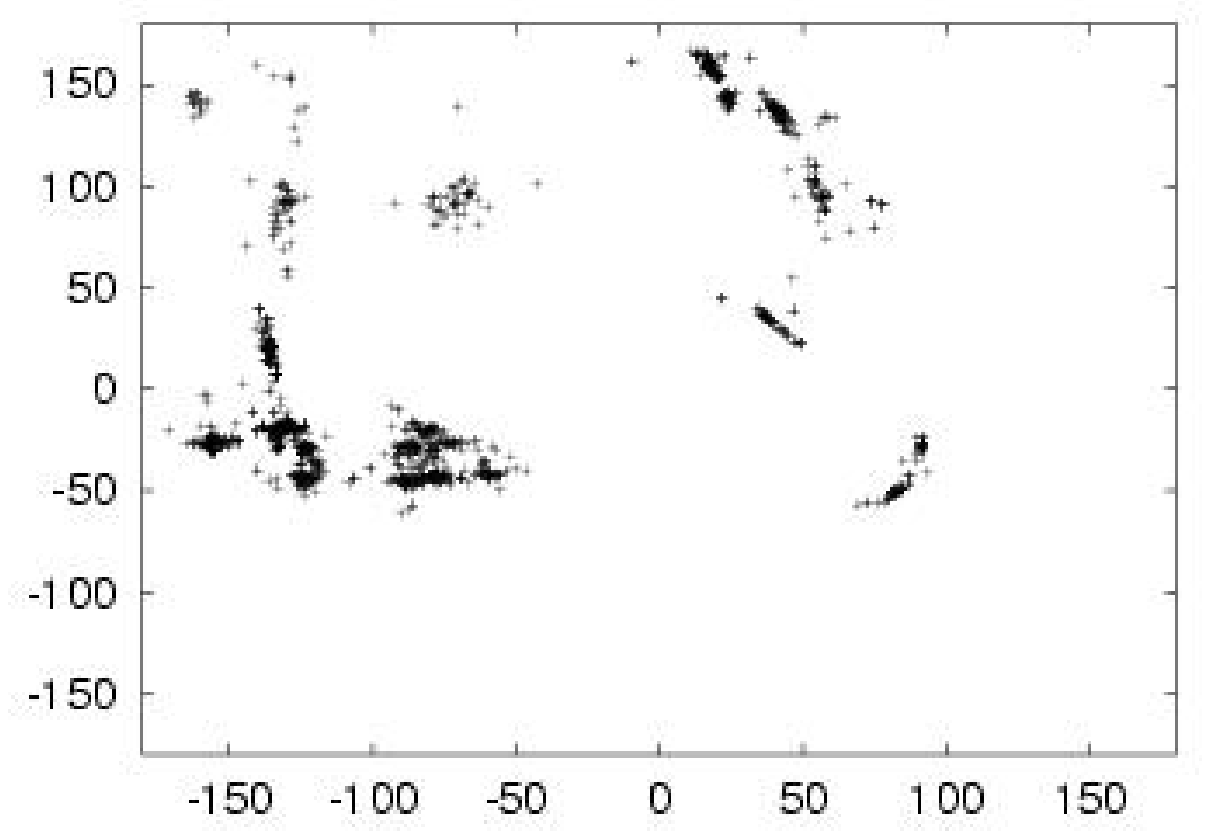,height=4.5cm,width=5cm} }}
\vspace*{-0.3cm}
\end{figure}
\begin{figure}[!h]
\centerline{\hbox{
\psfig{figure=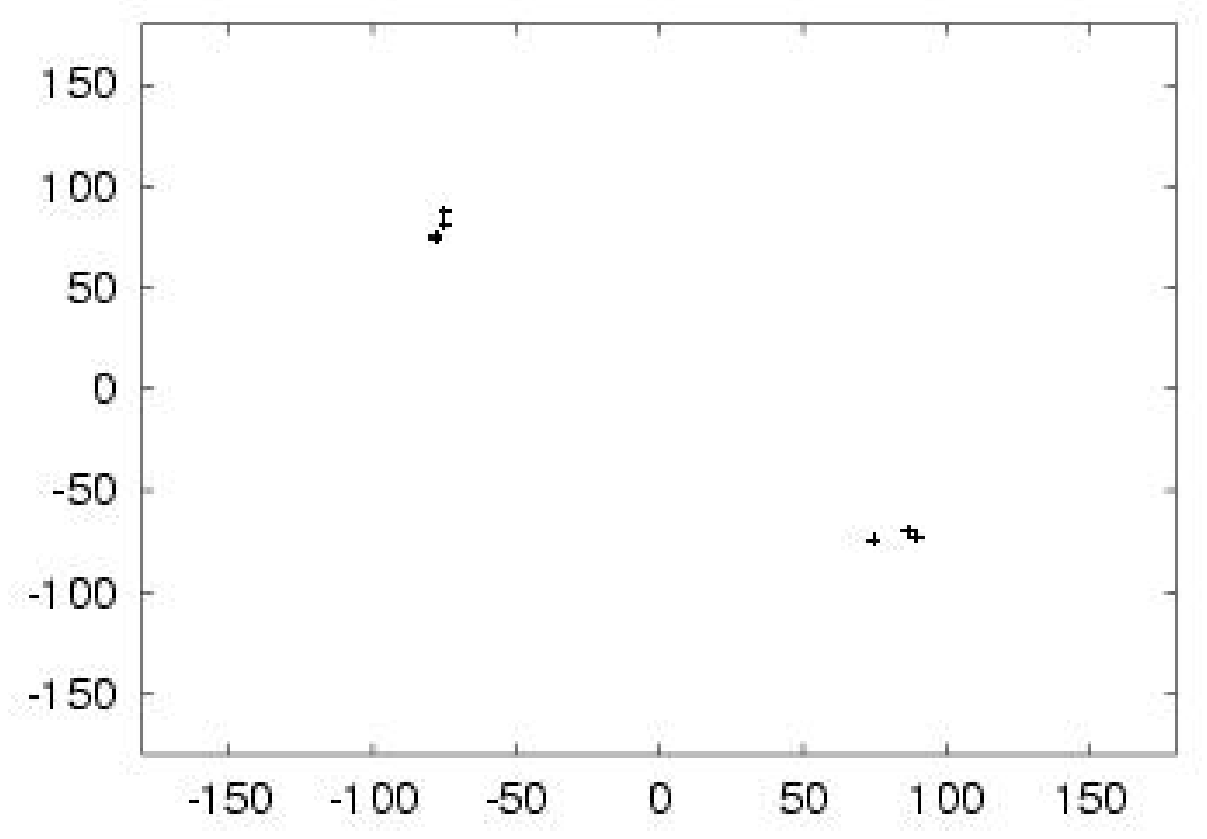,height=4.5cm,width=5cm}
\psfig{figure=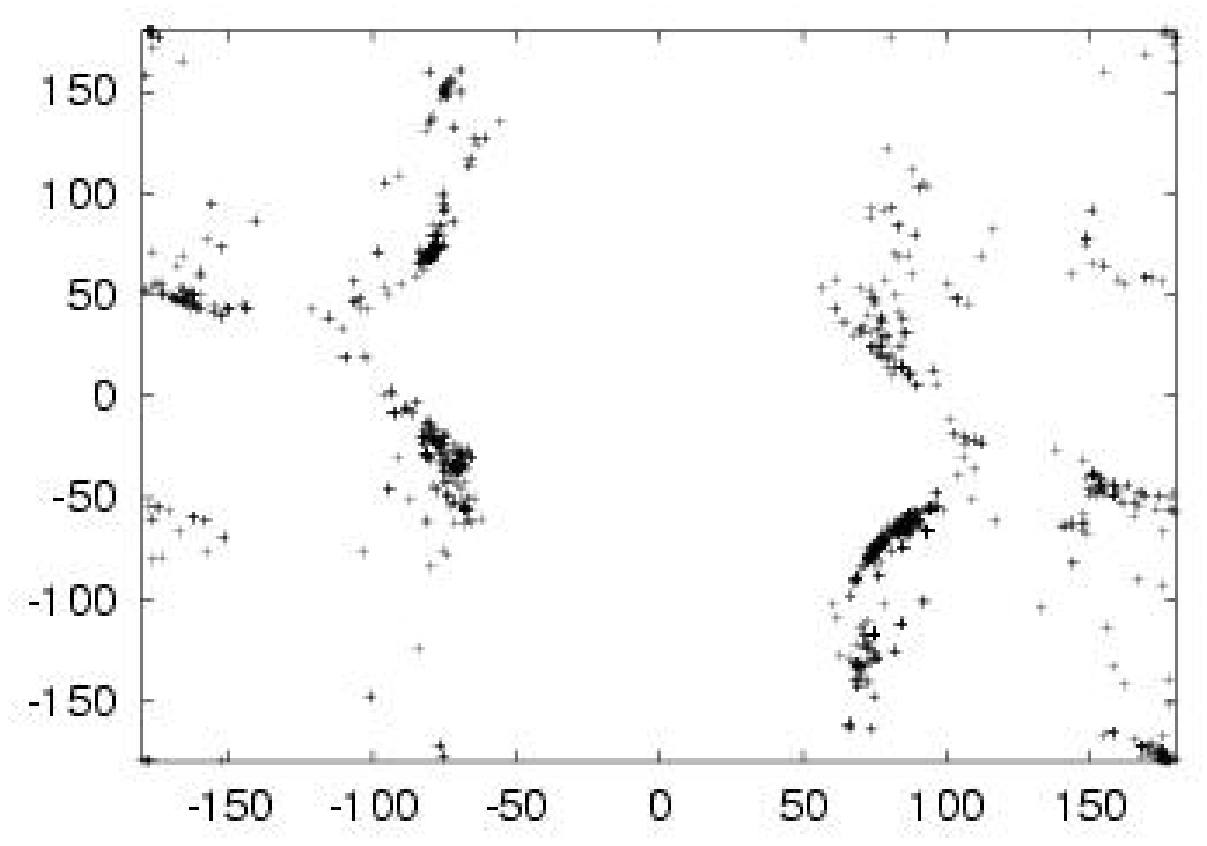,height=4.5cm,width=5cm} }}
\vspace*{-0.3cm}
\end{figure}
\begin{figure}[!h]
\centerline{\hbox{
\psfig{figure=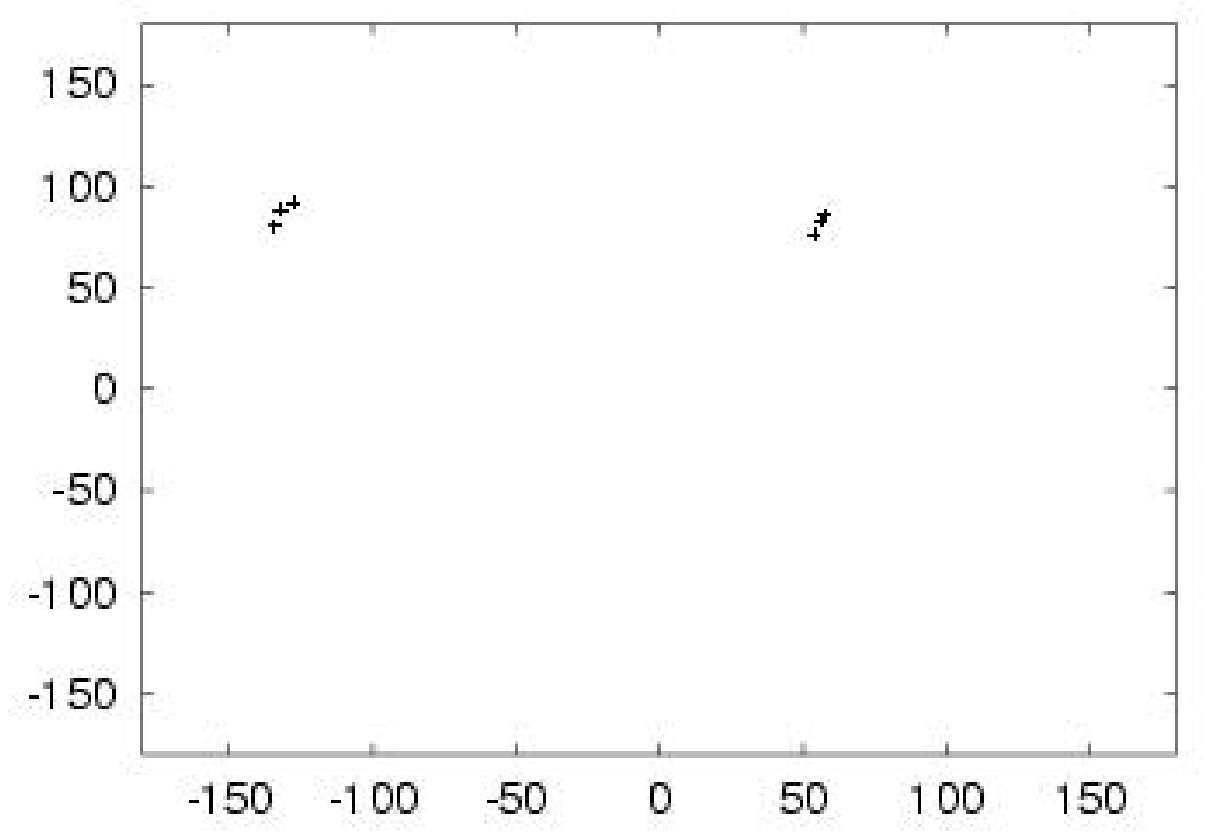,height=4.5cm,width=5cm}
\psfig{figure=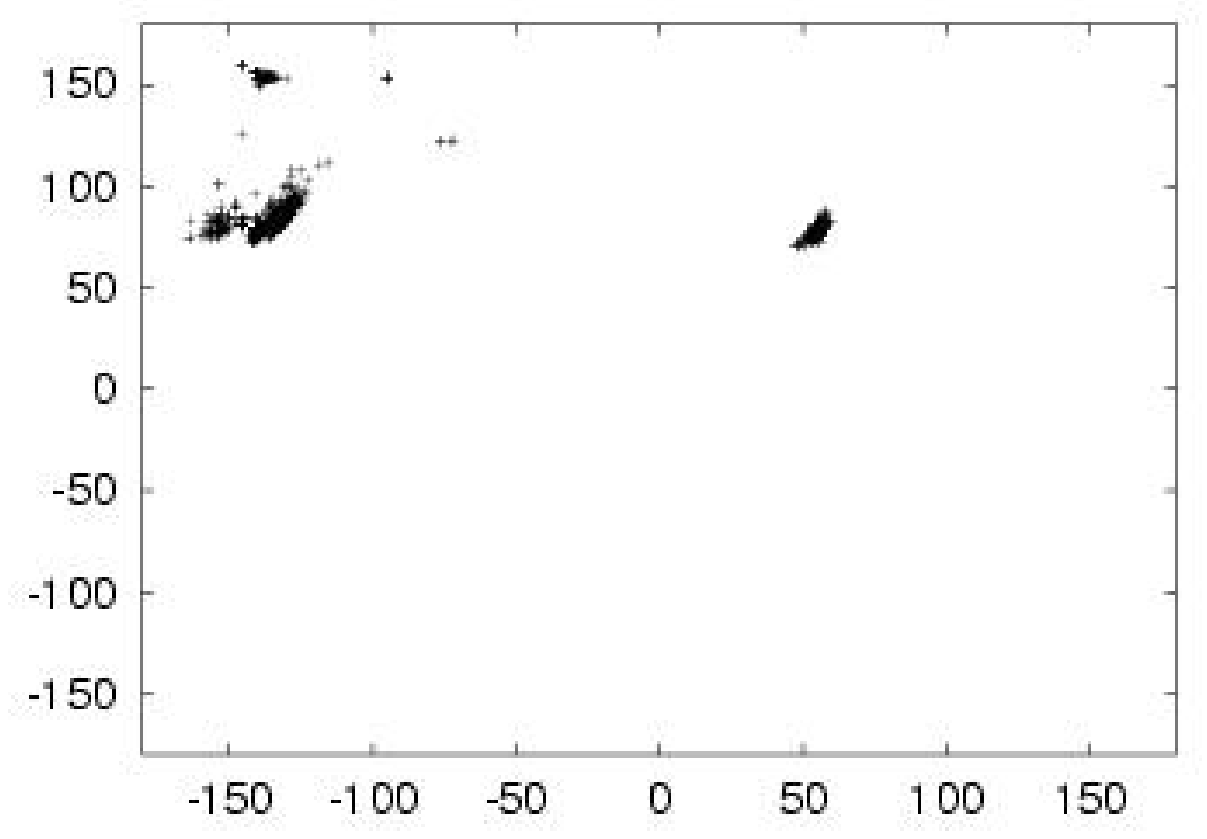,height=4.5cm,width=5cm} }}
\vspace*{-0.3cm}
\end{figure}
\begin{figure}[!h]
\centerline{\hbox{
\psfig{figure=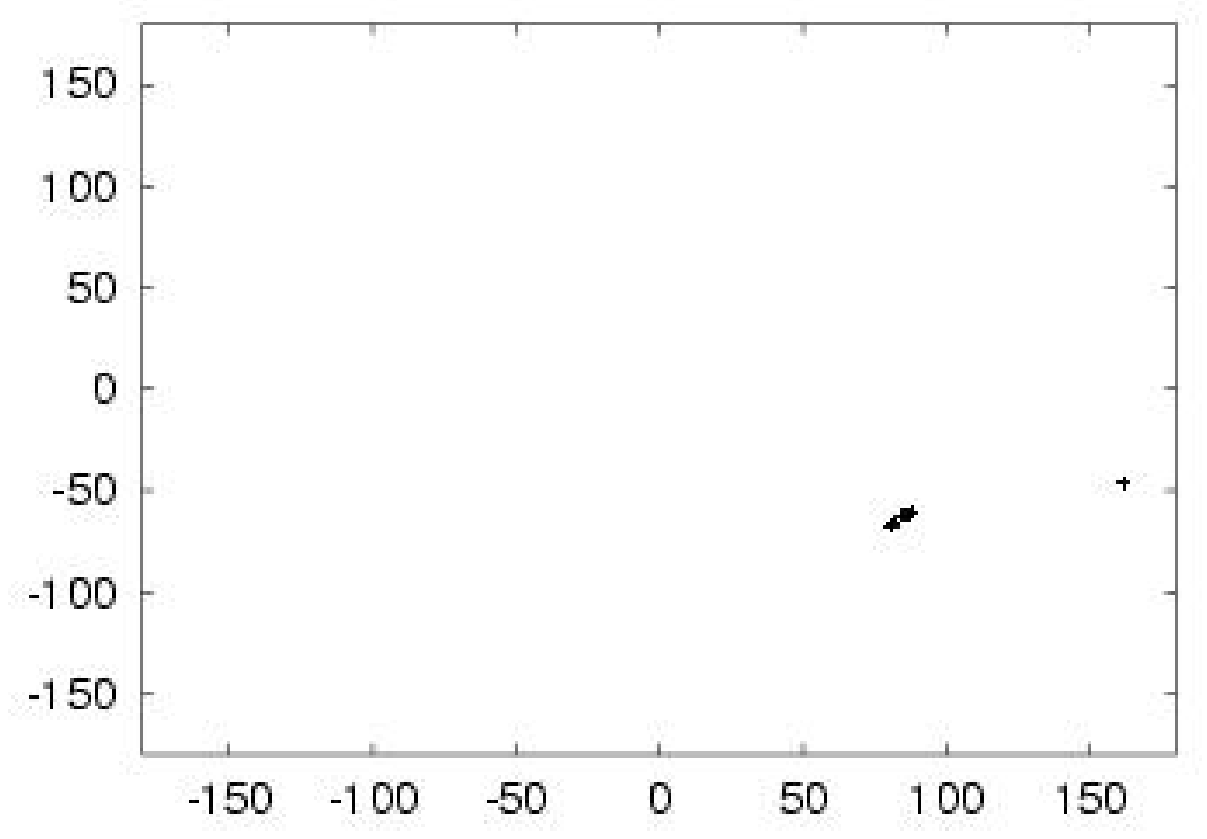,height=4.5cm,width=5cm}
\psfig{figure=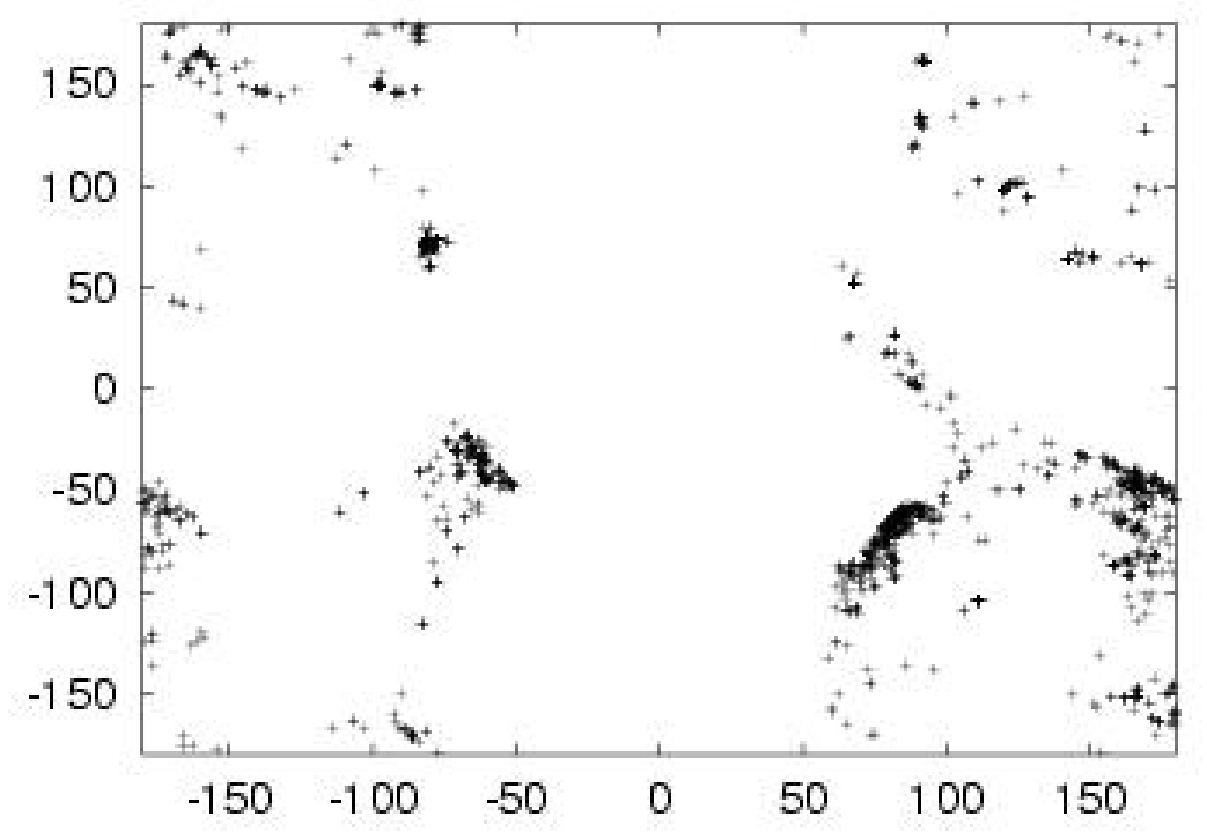,height=4.5cm,width=5cm} }}
\vspace*{-0.3cm}
\end{figure}
\begin{figure}[!h]
\centerline{\hbox{
\psfig{figure=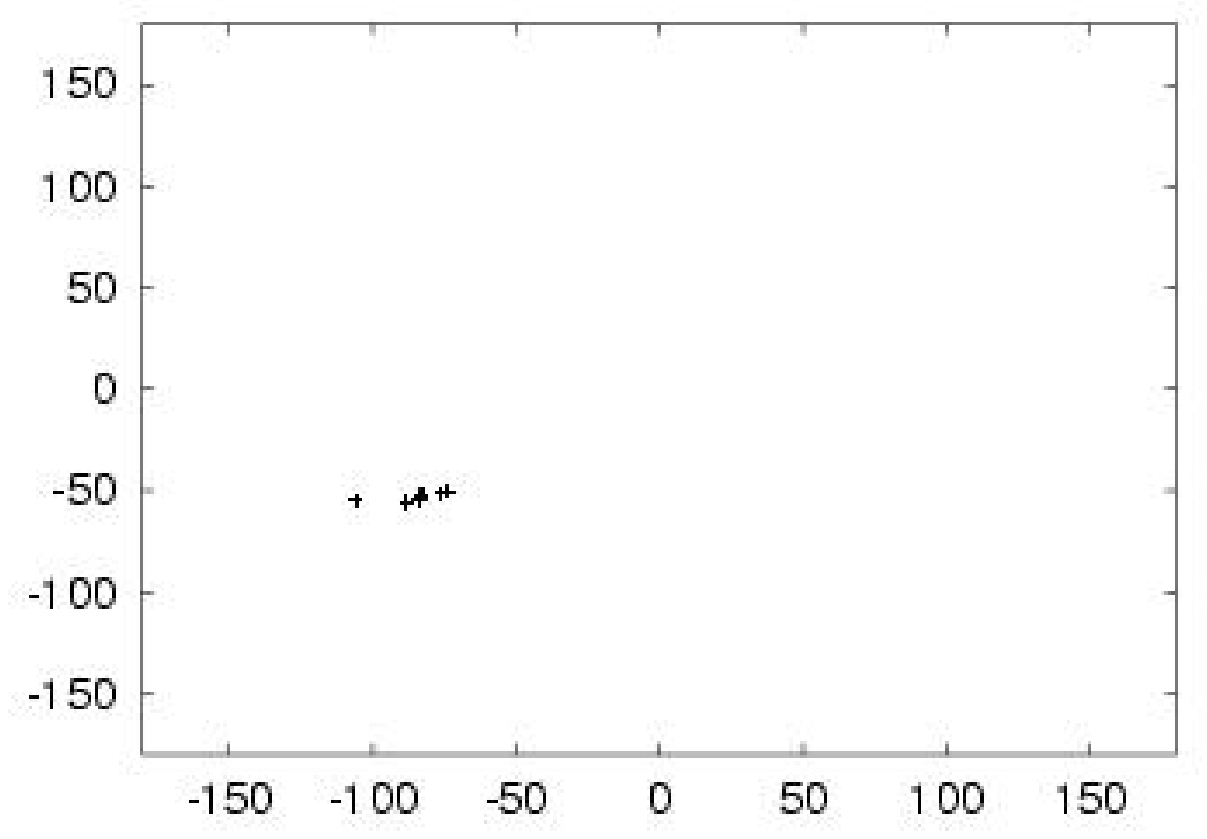,height=4.5cm,width=5cm}
\psfig{figure=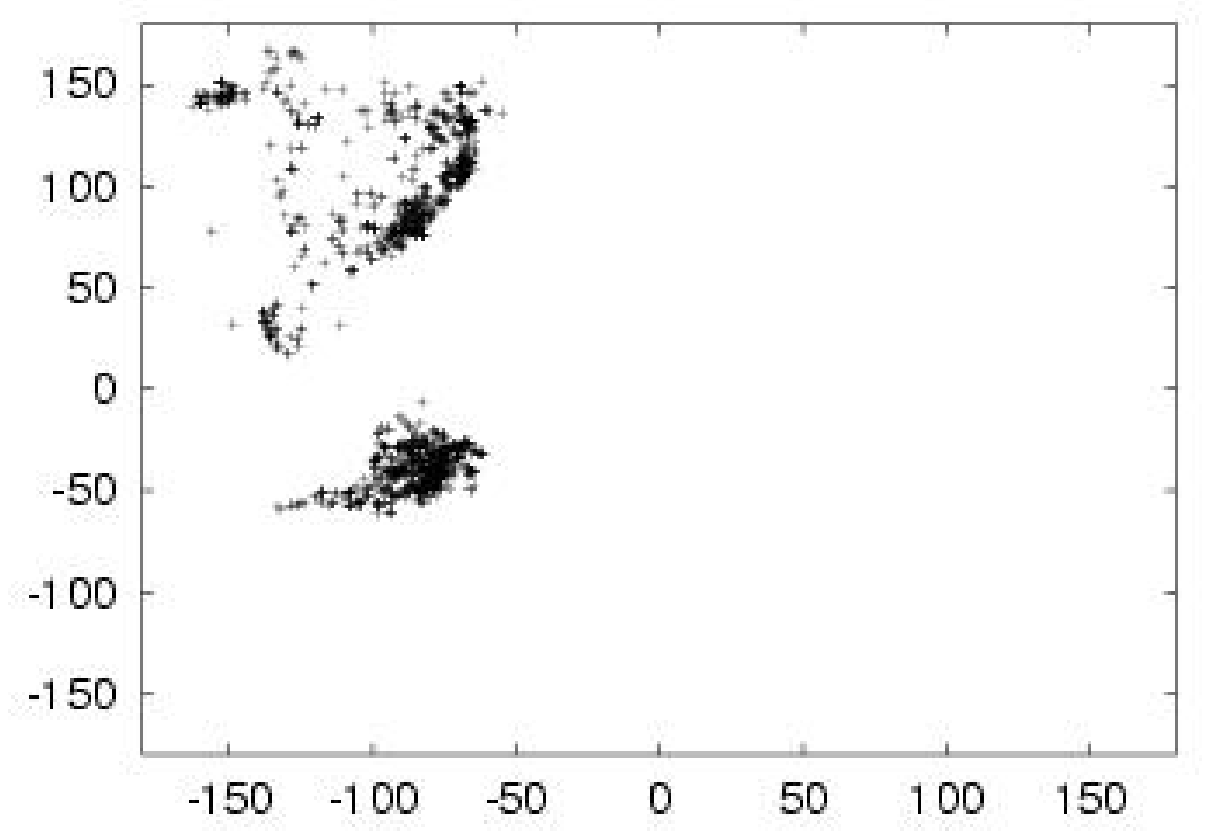,height=4.5cm,width=5cm} }}
\vspace*{-0.3cm}
\end{figure}
\begin{figure}[!h]
\centerline{\hbox{
\psfig{figure=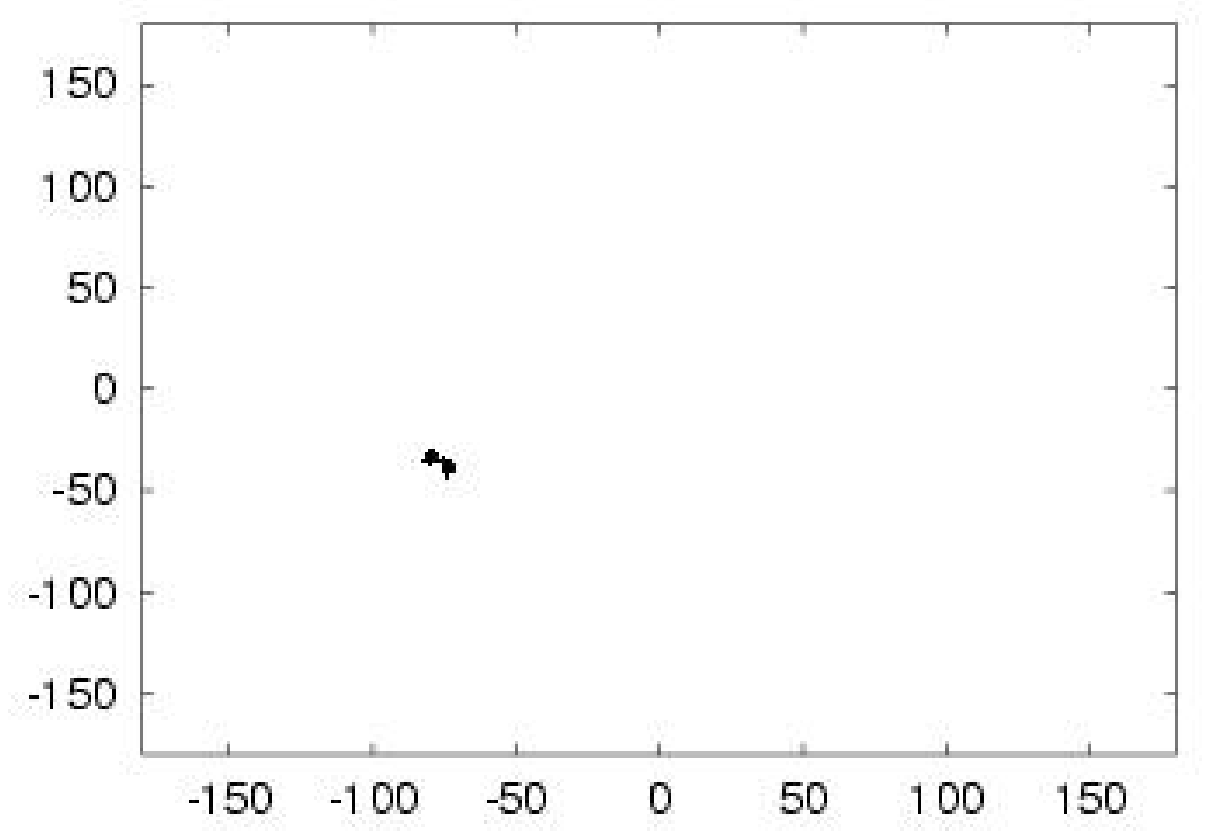,height=4.5cm,width=5cm}
\psfig{figure=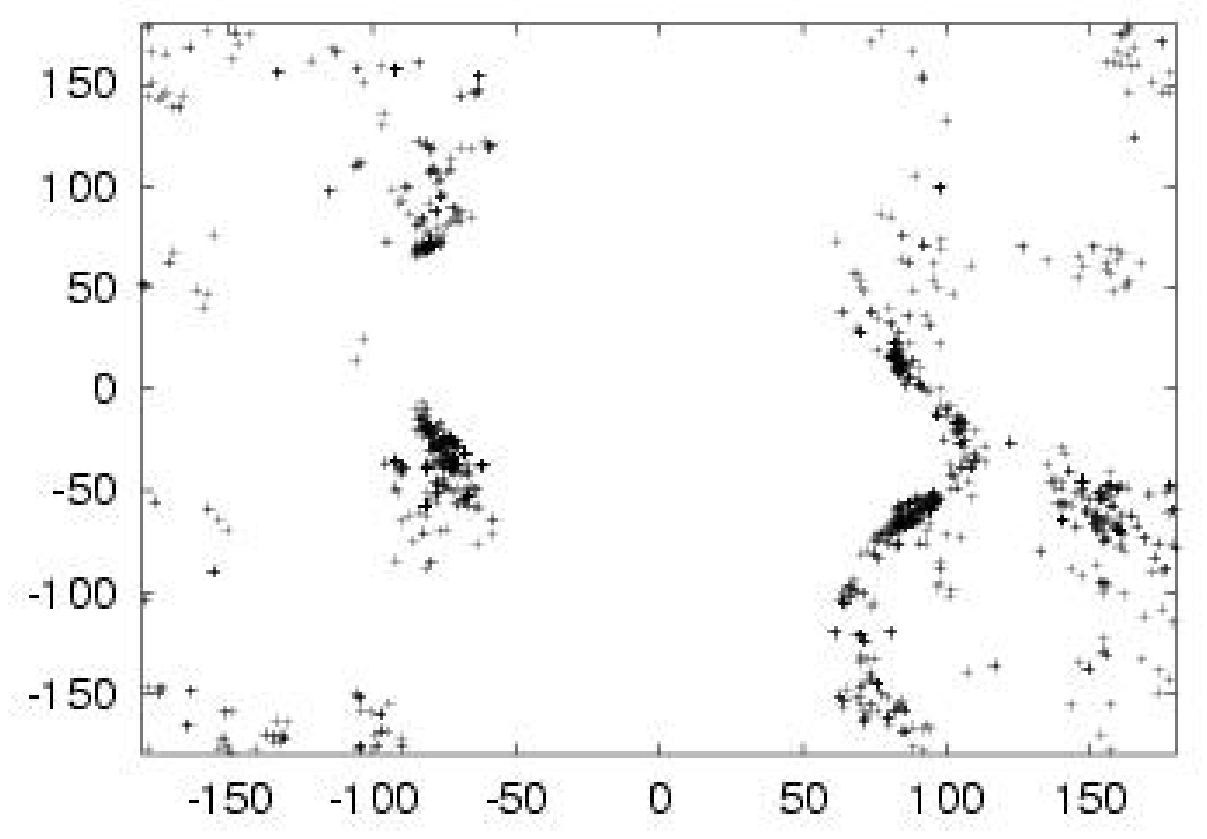,height=4.5cm,width=5cm} }}
\vspace*{-0.3cm}
\end{figure}

\clearpage
\pagebreak

\clearpage
\pagebreak
\begin{figure}[!t]
\psfig{figure=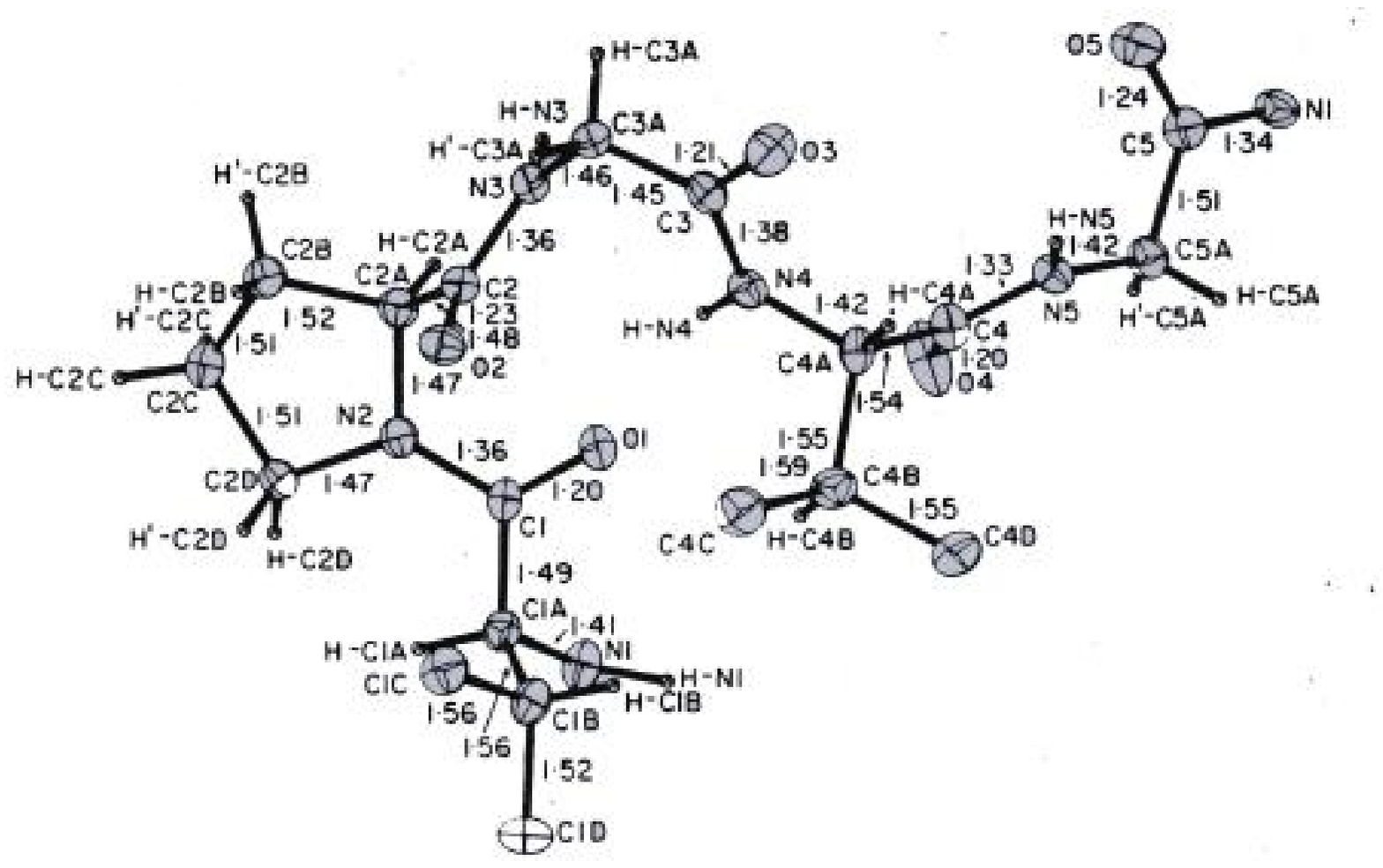,height=9.0cm,width=12cm}
\end{figure}

\clearpage

\pagebreak

\begin{table}
\caption{Number of Energy-Minimized Structures in Energy Bins
of $0.5$ kcal/mol above  $  E = - 17.94$ kcal/mol as obtained
by the MUCA and the MCM methods for the 
Val$^1$-Gly$^2$-Val$^3$-Pro$^4$-Gly$^5$-Val$^6$-Gly$^7$-Val$^8$-Pro$^9$ 
sequence.  The results of only $ 10^5$ sweeps are presented in the table.}
\label{xx}
\vspace{0.3cm}
{\centering \begin{tabular}{|c|c|c|c|}
\hline 
Bin (kcal/mol)&
Energy (kcal/mol)&
MUCA&
MCM (T=500K)\\
\hline 
\hline 
0.0 - 05&
-17.95 to -17.45&
1096&
1774\\
\hline 
0.5 - 1.0&
-17.45 to -16.95&
2374&
3462\\
\hline 
1.0 - 1.5&
-16.95 to -15.95&
2805&
4089\\
\hline 
1.5 - 2.0&
-15.95 to -15.45&
5038&
5902\\
\hline 
2.0 - 2.5&
-15.45 to -13.95&
4767&
5646\\
\hline 
2.5 - 3.0&
-13.95 to -13.45&
5117&
4764\\
\hline 
3.0 - 3.5&
-13.45 to -12.95&
5512&
3609\\
\hline 
3.5 - 4.0&
-12.95 to -12.45&
5116&
2213\\
\hline 
4.0 - 4.5&
-12.45 to -11.95&
4376&
2169\\
\hline 
4.5 - 5.0&
-11.95 to -11.45&
3632&
1275\\
\hline 
5.0 - 5.5&
-11.45 to -10.95&
2797&
1147\\
\hline 
5.5 - 6.0&
-10.95 to -10.45&
2470&
611\\
\hline 
\end{tabular}\par}
\vspace{0.3cm}
\end{table}

\bigskip

\end{document}